# Predicting synthesis window of *β*-TaON with thermodynamic modeling

**Dmitri LaBelle** and **Yong-Jie Hu**[†]

*Department of Material Science and Engineering, Drexel University*
*E-mail:* [†]yh593@drexel.edu,

**Abstract.** Phase-pure synthesis has been a major challenge for metal oxynitrides due to their sensitivity to synthesis conditions and the limited understanding of the underlying thermodynamics. The beta-phase tantalum oxynitride (β-TaON), a promising material for applications in photo- catalysis and energy storage, is particularly difficult to synthesize in a reproducible, phase-pure form. In this work, a computational thermodynamic model with experimental validation is presented to evaluate the phase-pure synthesis conditions for β-TaON via ammonolysis reactions. The finite-temperature thermochemical properties of the reactant, product, and byproduct phases are predicted via first-principles calculations with the quasi-harmonic approach (QHA) as well as implemented from available thermodynamic databases. With the thermochemical properties, a thermodynamic model based on the "CALculation of PHAse Diagrams" (CALPAHD) approach is developed to assess the phase equilibria associated with the synthesis reactions and correspondingly predict the synthesis window for β-TaON. A three-dimensional phase diagram is predicted as a function of gas composition and temperature, providing insights into optimal synthesis conditions. The computational predictions are further compared with available experimental data, offering a systematic framework for phase-pure β-TaON synthesis.

**Keywords:** Tantalum oxynitrides, Synthesis Window, First-principles Calculations, CALPHAD Modeling, Ammonolysis Reactions.

## 1. Introduction

Metal oxynitrides are a class of inorganic compounds characterized by a mixed anion sub-lattice containing both oxygen and nitrogen atoms. The mixing can occur in either a disordered or ordered configuration. Disordered mixing typically arises when nitrogen atoms, at doping levels, substitute for oxygen in the sublattice of oxide crystal structures [1–3]. As nitrogen content increases to concentrated levels, the system tends to favor long-range ordering, resulting in oxynitrides with distinct crystal structures that differ from their pure oxide or nitride counterparts [2].

Compared to oxygen, nitrogen has lower electronegativity and higher polarizability, which increases the covalency of metal-anion bonding in oxynitrides. Consequently, the 2p orbitals in oxynitrides form shallower electronic bands relative to those in oxides, leading to a reduced and tunable bandgap, depending on the oxygen-to-nitrogen ratio [4]. Moreover, the -3 valance state of nitrogen may allow the metal cations to achieve higher and mixed oxidation states compared to oxides. These combined effects of oxygen and nitrogen mixing in the same sublattice give oxynitrides a unique set of properties, making them promising for applications in photocatalysis, pigments, phosphors, dielectrics, and magnetic materials. For example, the reduced bandgap can shift the photocatalytical activity from the UV to the visible light range [5, 6]. In rare earth oxynitrides, the 5d orbital energy levels are lowered due to a stronger nephelauxetic effect and increased crystal field splitting, resulting in a red shift in emission wavelengths. Together with the additional effects such as symmetry distortions or anion ordering, it grants some oxynitrides interesting luminescence properties [7]. Additionally, for the late transition metals, the Fermi level in their (oxy)nitrides is usually at higher energies compared to corresponding oxides, making these compounds attractive as electrode materials for lithium or sodium ion batteries due to a decrease in the operating voltage [8].

Despite their remarkable properties, a major challenge for oxynitride materials is the control and reproducibility of their syntheses. Because of the high bond energy of the $N_2$ molecule and low electron affinity of the N atom, (oxy)nitrides generally show smaller energies of formation than oxides. As a result, although most oxynitrides are thermodynamically stable, the higher stability of oxides limits the synthesis window for oxynitrides, requiring precise control of temperature and oxygen activity to prevent overreaction or degradation [9]. The most common approach for oxynitride synthesis is via ammonolysis reactions, where oxide precursors react with flowing ammonia gas at moderate or high temperatures. Since the kinetics of ammonia decomposition at elevated temperatures is relatively slow, it essentially results in an extremely high nitrogen activity (or fugacity), allowing the nitrogen atoms to substitute oxygen in the anion sublattice, forming oxynitrides [2]. The exergonicity of ammonolysis reactions is highly sensitive to the reaction temperature and the activities of the oxygen and nitrogen species. Consequently, achieving phase-pure oxynitrides is a non-trivial task that requires meticulous optimization of key synthesis parameters, including temperature, gas flow rate, and reaction time.

Tantalum oxynitride (TaON) is among the oxynitrides that have proven challenging to syn- thesize in a controllable and reproducible way with high purity. The orthorhombic tantalum pen- toxide $Ta_2O_5$ with the p2mm space group is often used as the precursor phase. The synthesis reaction can be written as,

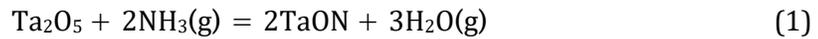
$$Ta_2O_5 + 2NH_3(g) = 2TaON + 3H_2O(g) \qquad (1)$$

To enable phase-pure synthesis of $\beta$-TaON, it is crucial to precisely control the activities of nitro- gen and oxygen species. For instance, excessively high nitrogen activity or low oxygen activity can lead to the formation of the $Ta_3N_5$ byproduct via a reaction,

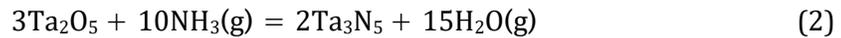
$$3Ta_2O_5 + 10NH_3(g) = 2Ta_3N_5 + 15H_2O(g) \qquad (2)$$

Conversely, insufficient nitrogen activity or excessive oxygen activity would prevent the am- monolysis of the oxide precursor. Therefore, in the practical synthesis, the ammonia gas often flows together with a certain amount of $H_2$ and $H_2O$ gases, allowing the nitrogen and oxygen activ- ities to be controlled by adjusting the gas composition as well as reaction temperatures. However, due to a lack of understanding of the reaction thermodynamics, the previous attempts for optimal gas composition and temperature were mainly carried out based on trial-and-error. As a result, the recipes reported in the literature by individual groups showed certain discrepancies [10–13], which could be difficult to reproduce in other laboratories due to small differences in the experimental setups. Additionally, the TaON composition possesses multiple polymorphic crystal structures, of which the formation was also sensitive to the synthesis conditions [2, 14, 15].

Among the various polymorphs of tantalum oxynitride, $\beta$-TaON is the most stable under ambient conditions and has been the most extensively studied. Although reports on the successful synthesis of $\beta$-TaON have been prevalent in the literature, only two prior studies, to the authors' knowledge, have systematically investigated its synthesis window [10, 11]. In 1972, Swisher and Read performed ammonolysis reactions with $Ta_2O_5$ at a constant temperature of 1100K and various ratios between the partial pressures of $NH_3$ and $H_2$, $p_{NH3}/p_{H2}^{3/2}$, and the partial pressure of $H_2O$ and $H_2$, $p_{H2O}/p_{H2}$ [10]. Based on the reaction products, they manually derived a phase diagram on the axes of $p_{NH3}/p_{H2}^{3/2}$ and $p_{H2O}/p_{H2}$, indicating two-phase equilibrium boundaries between $Ta_2O_5$ and $\beta$-TaON, and between $\beta$-TaON and $Ta_3N_5$. The synthesis window for $\beta$-TaON was defined as the region between these phase boundaries, which have been widely referenced in later studies. In 2015, de Respinis et al. [11] conducted a comprehensive set of experiments at the same synthesis temperature but found that $\beta$-TaON formed at a much lower ratio of $p_{NH3}/p_{H2}^{3/2}$, falling within the single-phase region of $Ta_2O_5$ as proposed by Swisher and Read. Furthermore, under the $p_{NH3}/p_{H2}^{3/2}$ and $p_{H2O}/p_{H2}$ conditions identified by Swisher and Read as single phase for $\beta$-TaON, de Respinis et al. observed the formation of a mixture of $\beta$-TaON and $Ta_3N_5$ [11]. These discrepancies in the reported synthesis windows for $\beta$-TaON underscore the need for a deeper in- vestigation into the thermodynamics and phase equilibrium associated with the synthesis reactions to enable a more systematic and accurate prediction of the synthesis window.

In this work, we present a computational study for a prediction of the synthesis window for $\beta$-TaON via ammonolysis reaction regarding the composition ratios of the reactive gas as well as the reaction temperature. First, we utilize DFT calculations to predict the thermochemical properties of $\beta$-TaON and $Ta_3N_5$ as a function of temperature, including heat capacity, enthalpy, entropy, and Gibbs free energy. To capture the entropic contributions at finite temperatures, the quasi-harmonic approach is employed in coupling with first-principles phonon calculations. Second, we develop a thermodynamic model by means of CALculation of PhAase Diagram (CALPHAD) method to assess the phase equilibria among $Ta_2O_5$, $\beta$-TaON, and $Ta_3N_5$, which are the precursor, synthesis target, and over-reacted byproduct phases, respectively. The model also implements the thermochemical properties of the gaseous phase and $Ta_2O_5$ from the Scientific Group Thermodata Europe (SGTE) Substances database version 6 (SSUB6) [16]. Based on the thermodynamic assessment, a three-dimensional (3D) synthesis window for $\beta$-TaON is outlined as a function of $p_{NH3}/p_{H2}^{3/2}$, $p_{H2O}/p_{H2}$, and temperature. Finally, we validate our computational predictions by comparing them with available experimental data in the literature.

## 2. Methods

### 2.1. First-principles calculations

First-principals calculations were carried out based on DFT using projector augmented wave (PAW) method [17,18], as implemented in the Vienna Ab initio Simulation Package (VASP) [19]. The generalized gradient approximation (GGA) functional developed by Perdew-Burke-Ernzerhof (PBE) [20] was chosen to describe the exchange-correlation interactions. The crystal structure of the $\beta$-TaON molecule was sourced from the Materials Project [21] for TaNO (mp-4165) from database version v2023.11.1. The energy convergence criterion of the electronic self-consistency was set as 10⁻⁶ eV and 10⁻⁸ eV for the relaxation and electronic structure calculation, respec- tively. The Brillouin-zone integration of electronic structure was performed using a tetrahedron method with Blöchl corrections, while Gaussian-smearing was used for structure relaxation. The width of the smearing was chosen as 0.05 eV. The energy cutoff on the wave function was taken as 520 eV. A 20×20×20 k-point Γ-centered mesh was used for relaxation calculations. The phonon calculations were carried out using the density functional perturbation theory (DFPT) method, where a 3 × 3 × 3 supercell expansion of the base $\beta$-TaON was used for the phonon calculations. Phonon calculations were post-processed using Phonopy [22] code.

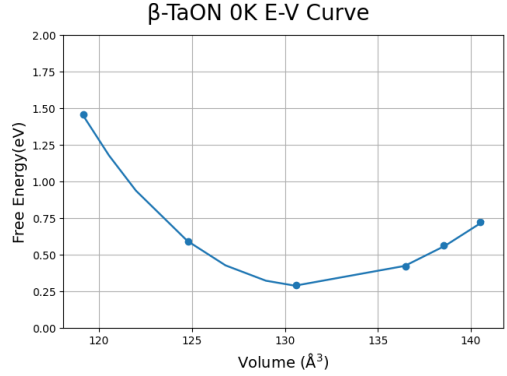

**Fig. 1.** Static energy vs. volume curve for $\beta$-TaON at 0K

*2.2. Prediction of finite temperature thermodynamic properties by quasi-harmonic approximation*

Based on the QHA, Helmholtz free energy of a stoichiometric solid phase could be expressed as a function of volume and temperature,

$$F(V, T) = E(V) + F_{vib}(V, T) + F_{el}(V, T) \tag{3}$$

where $E$ is the static energy at 0K, $F_{vib}$ is the vibrational free energy, and $F_{el}$ is the contribution due to thermal excitation of electrons [23, 24]. The total static energy at 0K is obtained by fitting the energy-volume data obtained by DFT calculations at different volumes to a 4-parameter Birch- Murnaghan equation of state (EOS) [25], as shown in figure 1:

$$E(V) = a + bV^{-2/3} + cV^{-4/3} + dV^{-2} \tag{4}$$

where $a, b, c$, and $d$ are fitting coefficients from which the physical parameters, including $E_0$ equi- librium static energy at 0K, $V_0$ volume, $B_0$ bulk modulus, and $B'$ its first derivative with respect to
pressure, can be derived. The $F_{vib}$ of $\beta$-TaON was obtained though the phonon model scheme from the phonon den- sity of state (DOS) at a given volume $V$ is discribed by the following:

$$F_{vib}(V,T) = k_B T \int_0^\infty \ln\left[2\sinh\frac{\hbar\omega}{2k_B T}\right] g(\omega)d\omega, \tag{5}$$

where $k_B$ is Boltzmann's constant, $\hbar$ the reduced Planck constant, and $g(\omega)$ phonon DOS, with $\omega$ being phonon's natural frequency [24]. The $F_{el}$ was computed from the electron density of

states (DOS) as follows

$$F_{\text{vib}}(V,T) = \frac{9}{8}k_B\theta_D + k_BT\left\{3\ln\left[1-\exp\left(-\frac{\theta_D}{T}\right)\right] - D\left(\frac{\theta_D}{T}\right)\right\}, \tag{6}$$

where $n(\epsilon)$ is the electronic DOS, $\epsilon$ the energy eigenvalue, $\epsilon_F$ Fermi level energy, and $f$ the Fermi-Dirac distribution function as electrons are fermions. The obtained $F$ at zero pressure is taken as $G$ due to the negligible $PV$ contribution for solid phases at the ambient pressure. Other thermo-dynamic properties, such as entropy $S$, enthalpy $H$, and heat capacity at constant pressure $C_P$, are computed from $G(V, T)$ using fundamental thermodynamic relations.

*2.3. CALPHAD Modelling*

The Gibbs free energy function of studied β-TaON at room temperature has been described as [5, 26, 27]:

$$G(T) - H^{SER} = a + bT + cT\ln T + dT^2 + eT^{-1} + fT^3 \tag{7}$$

where $a, b, c, d, e$, and $f$ are the model parameters evaluated from the thermochemical data calcu-lated by the first-principles QHA method described above. $H^{SER}$, refereed to as the stable element reference (SER) state, represents the enthalpies of the pure elements forming β-TaON at 298.15 K and 1 bar in their stable states. Such an exercise also needed to be done with the expected ni-trite by-product $Ta_3N_5$. For the precursor oxide state, $Ta_2O_5$, the Gibbs free energy functions was available and adopted from a commercial database SSUB6 as implemented in the Thermo-Calc software [16], with additional support referencing the model parameters for the exact precursor β-$Ta_2O_5$ polymorph [28]. While reacting gasses and carriers like $H_2O$, $NH_3$, and $H_2$ were readily available in the Thermo-Calc database. The Gibbs free energy functions in the SSUB6 database were derived from high-quality experimental thermochemical data, such as finite temperature en-thalpy and entropy of formation, and heat capacity, and were thus considered to be reliable and were widely used to model phase stability and transformation involving gas molecules [29, 30]. The modeled Gibbs free energy function was then used to calculate the expected phase transition lines as function of activity coefficients of pumped in reactive gasses at equilibrium using global free energy minimization approach implemented in the Thermo-Calc software.

We choose to define the x and y axis of our phase diagram as analogous to the 1972 Swisher and Read paper [10], where they've defined as

$$x = \frac{p_{H_2O}}{p_{H_2}}$$

and

$$y = \frac{p_{NH_3}}{p_{H_2}^{3/2}}$$

as our axes as shown in figure 2.

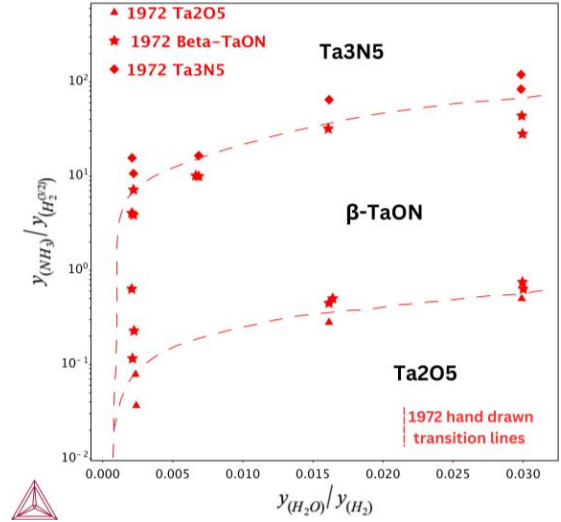

**Fig. 2.** A hand-drawn phase diagram for the synthesis window of $\beta$-TaON at 1100K, which is adapted from 1972 Swisher and Read [10].

What is important to note is that this pressure fraction is better understood as concentration coefficient of reacting gasses as this reaction occurs at ambient pressure using an argon carrier and assuming no argon decomposition. Thereby we correctly define our axes;

$$x = \frac{y_{H_2O}}{y_{H_2}}$$

and

$$y = \frac{y_{NH_3}}{y_{H_2}^{3/2}}$$

The initial justification for this axes definition is Swisher and Read's original interpretation of this system as an ion Exchange Reaction – based on the ratio of the oxygen reaction to the nitrogen reaction and the resulting ratio of N/O. This then necessitates the use of reaction equilib- rium constants based on the basic decomposition of $NH_3$ and $H_2O$ at the same temperature as our working model defined exactly below:

$$NH_3 \rightarrow \frac{1}{2}N_2 + \frac{3}{2}H_2; K1 = \frac{y(H_2)^{\frac{3}{2}}}{y(NH_3)} y(N_2)^{\frac{1}{2}} \quad (8)$$

$$H_2O \rightarrow \frac{1}{2}O_2 + H_2; \quad K2 = \frac{y(H_2)}{y(H_2O)} y(O_2)^{\frac{1}{2}} \quad (9)$$

The values of $K1$ and $K2$ we want to use are obtained by setting our conditions as T = 1100K and P = 1atm in Thermo-Calc and computing for the water and ammonia equilibrium respectively across a temperature range. This will be useful in future studies when we want to attempt this experiment at different temperatures as that will necessitate different $K1$ and $K2$ val- ues. By digitizing the data values from the 1972 phase diagram (figure 2), we've been able to formally compute the optimal phase transition lines given the DFT G(T) approximation combined optimized to the '72 data using thermocalc software. Where we take out optimized G(T) terms for our three crystals, $Ta_3N_5$, TaON, and $Ta_2O_5$, and tack on an additional energy and entropy optimizing variable to $Ta_3N_5$ and TaON. We omit $Ta_2O_5$ from this additional optimization as we have a most confident handle on the oxide Gibbs free energy function. From optimization we were able to obtain new modeling results.

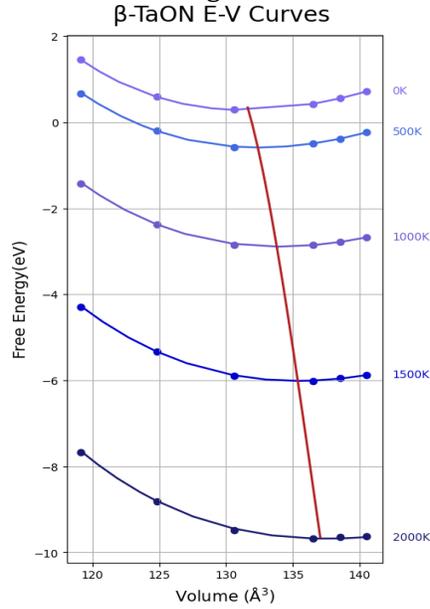

**Fig. 3.** Quasi-Harmonic approximation analysis of $\beta$-TaON from 0-2000K.

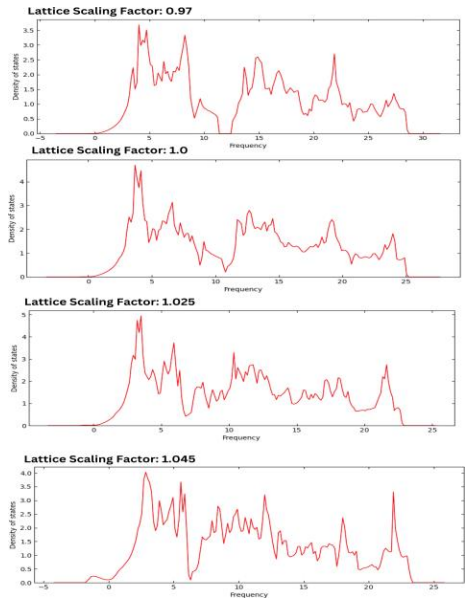

**Fig. 4.** Phonon density of state plots for a sampling of lattice spacing factors used in QHA

analysis. Note the imaginary phonon modes present for 1.045, which implies instability at higher temperatures.

## 3. Results and discussion

### 3.1. Thermochemical properties of β-TaON

From DFT PBE based calculations we were able to preform a 0K QHA analysis with results shown in figure 1. By the use of the programming package Phonopy [22] we were able to combine our 0K E-V curve with thermal property data for a sampled volume range utilizing a lattice constants: [0.97, 0.985, 1, 1.015, 1.02, 1.025] to stretch and squeeze our unit cell. The resulting free energy volume curves for temperatures ranging from 0 to 2000K are shown in figure 3.

The irregular lattice scaling factor spacing was motivated by the imaginary phonon mode results which came about from larger lattice spacing analysis, with prior lattice spacing values of 1.03 and 1.045 yielding a cascading error when applied to phonon analysis. Figure 4 includes phonon density of state graphs for a sampling of lattice scaling factors which highlights this point. We'll note the bump lower than our x angle 0, which signifies instability at this volume. This is to be expected as the $\beta$-TaON phase is considered to be the low-temperature stable phase of the TaON family of crystals.

From phonon analysis we are able to derive a number of important thermochemical properties for both experimental verification of the model as well as going on to form the basis Gibbs free energy function for optimization. The bulk modulus as well as the unsifted Gibbs free energy, the heat capacity, and the entropy of $\beta$-TaON is shown in figures 5 and 6. As explained in the CALPHAD modeling section, we combine these values of Gibbs and entropy with experimen- tally verified enthalpy values to form a stable element reference enthalpy state. We now have a complete version of equation 7, ready for thermochemical optimization in CALPHAD.

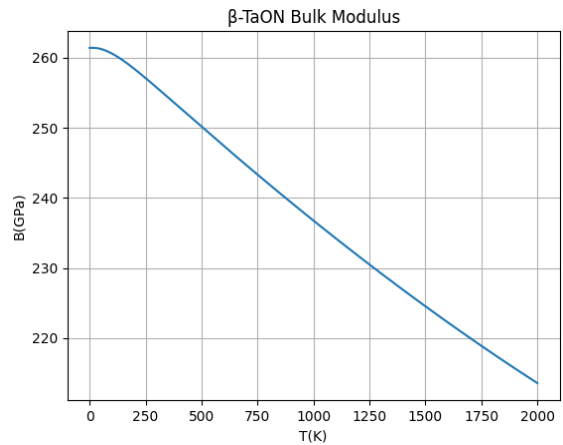

**Fig. 5.** Bulk modulus of $\beta$-TaON from 0-2000K

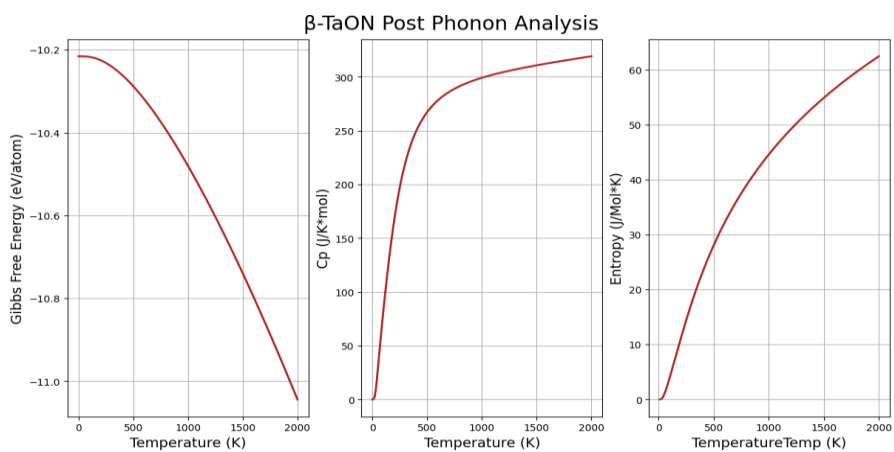

**Fig. 6.** Gibbs Free energy, Heat capacity, and entropy as a function of temperature for $\beta$-TaON from 0-2000K

## 3.2. Thermodynamic modeling of the synthesis window

While the digitized data and model made by the 1972 paper informed our initial optimization synthesis window calculation, this was quickly abandoned in favor of newer data from the 2015 $\beta$-TaON paper [11], as well as our own experimental runs.

The initial effort into basing our model on the 1972 data forced radical destabilization of the TaON and $Ta_2N_5$ phases. This occurred on a scale of demanding approximately an extra 5134 J/mol-atom for the TaON phase and 11054 J/mol-atom for $Ta_2N_5$. This large destabilization value as well as the nonphysical exponential decay of the original phase change lines as shown in figure 2 led us to seeking out other experimental sources of data, which lead us to utilizing the 2015 paper. Utilizing the 2015 $Ta_2O_5$, TaON+$Ta_3N_5$ mixed, and $Ta_3N_5$ data to guide our optimization model, we were able to obtain reasonable destabilization. Utilizing the Gibbs phase rule, it's vital to realize it is nonphysical to have a mixed phase region in our phase diagram. Therefore in optimization the TaON + $Ta_3N_5$ data point was treated as a pure TaON point. With lower destabilization values of approximately 4093 J/mol-atom for TaON and 3071 J/mol-atom for $Ta_3N_5$, the fitted 2015 data is shown in comparison to the 1972 data in figure 7.

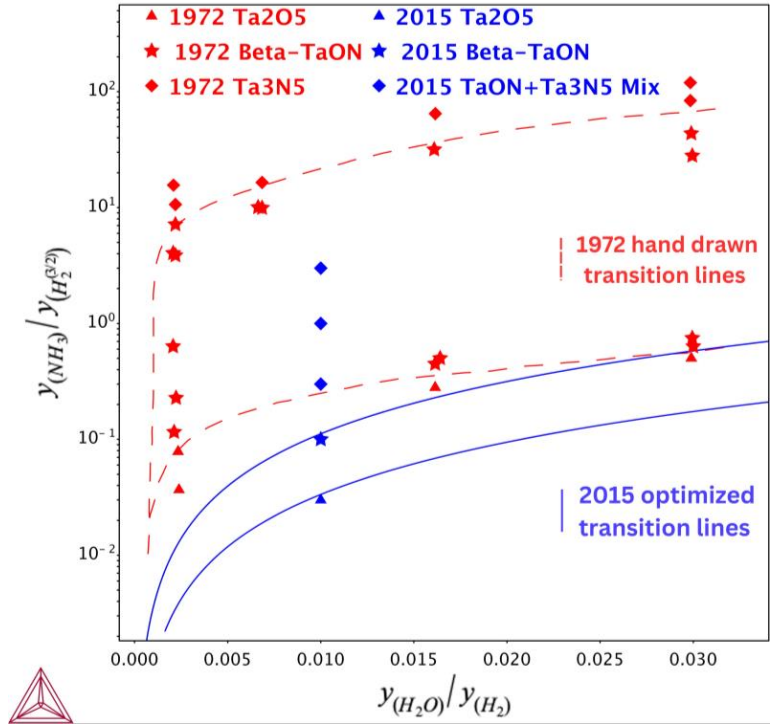

**Fig. 7.** The phase transition boundaries predicted from the optimized thermodynamic model (solid blue line) in comparison with previous experimental attempts and hand-drawn transition boundaries.

## 4. Conclusion

In this work, we developed a thermodynamic model using a combination of first-principles calculations and the CALPHAD method to assess and predict reaction thermodynamics and phase equilibria involved in the ammonolysis-based synthesis of $\beta$-TaON. The finite-temperature thermochemical properties of the $\beta$-TaON and $Ta_3N_5$ phases were predicted using the quasi-harmonic approximation approach, incorporating first-principles phonon calculations to account for entropic contributions. The thermodynamic model took the thermochemical properties and available Gibbs free energy functions of the gas and $Ta_2O_5$ phases in the SSUB database as inputs and predicted a phase diagram detailing the equilibrium among $Ta_2O_5$ (precursor), $\beta$-TaON (synthesis target), and $Ta_3N_5$ (over-reacted byproduct) phases as a function of temperature and oxygen/nitrogen activities. These activities were further interpreted in terms of reactive gas composition ratios, specifically $y_{NH_3}/y_{H_2}^{3/2}$ and $y_{H_2O}/y_{H_2}$, to map out a 3D synthesis window for $\beta$-TaON. The computational predictions are further compared and validated with available experimental data in the literature.